\documentclass[preprint,showpacs,showkeys,amsmath,amssymb,superscriptaddress,nofootinbib,floatfix]{revtex4}
\usepackage{graphicx}
\usepackage{bm}

\newlength{\dummysp}
\newcommand{\diag}{\mathop{{\hbox{diag} \, }}\nolimits}

\newcommand{\real}{\mathop{{\hbox{Re} \, }}\nolimits}

\newcommand{\half}{\frac{1}{2}}

\newcommand{\beq}{\begin{eqnarray}}
\newcommand{\eeq}{\end{eqnarray}}
\newcommand{\nnn}{ \nonumber \\ }
\newcommand{\p}{{\partial}}

\newcommand{\chib}{{\bar \chi}}

\newcommand{\e}{{\epsilon}}
\newcommand{\s}{{\sigma}}
\newcommand{\vev}[1]{{\langle #1 \rangle}}

\newcommand{\ord}[1]{{{\cal O}(#1)}}
\newcommand{\gappeq}{\mathrel{\rlap {\raise.5ex\hbox{$>$}}
{\lower.5ex\hbox{$\sim$}}}}
\newcommand{\lappeq}{\mathrel{\rlap{\raise.5ex\hbox{$<$}}
{\lower.5ex\hbox{$\sim$}}}}
\newcommand{\myref}[1]{(\ref{#1})}

\newcommand{\ben}{\begin{enumerate}}
\newcommand{\een}{\end{enumerate}}

\newcommand{\psib}{{\bar \psi}}

\newcommand{\ddd}{\nnn &&}

\newcommand{\bit}{\begin{itemize}}
\newcommand{\eit}{\end{itemize}}

\def\[{\left [}
\def\]{\right ]}
\def\({\left (}
\def\){\right )}

\def\nott#1{\setbox0=\hbox{$#1$}                
   \dimen0=\wd0                                 
   \setbox1=\hbox{/} \dimen1=\wd1               
   \ifdim\dimen0>\dimen1                        
      \rlap{\hbox to \dimen0{\hfil/\hfil}}      
      #1                                        
   \else                                        
      \rlap{\hbox to \dimen1{\hfil$#1$\hfil}}   
      /                                         
   \fi}                                         %

\begin{document}


\renewcommand{\thefootnote}{\fnsymbol{footnote}}

\author{Joel Giedt\footnote{{\tt giedtj@rpi.edu}} }
\affiliation{Department of Physics, Applied Physics, and Astronomy,
    Rensselaer Polytechnic Institute,
    110 Eighth Street, Troy, New York 12180-3590 USA}

\author{Andrew Skinner\footnote{{\tt askinner@skidmore.edu}} }
\affiliation{Physics Department, Skidmore College, 815 North Broadway, Saratoga Springs, NY 12866 USA}
\affiliation{Department of Physics, Applied Physics, and Astronomy,
    Rensselaer Polytechnic Institute,
    110 Eighth Street, Troy, New York 12180-3590 USA}

\author{Saroj Nayak\footnote{{\tt nayaks@rpi.edu}} }
\affiliation{Department of Physics, Applied Physics, and Astronomy,
    Rensselaer Polytechnic Institute,
    110 Eighth Street, Troy, New York 12180-3590 USA}

\title{Effects of flavor-symmetry violation from staggered fermion lattice
simulations of graphene}

\begin{abstract}
We analyze the effects of flavor splitting from staggered fermion lattice
simulations of a low energy effective theory for graphene.  Both the unimproved action and the
tadpole improved action with a Naik term show significant
flavor symmetry breaking in the spectrum of
the Dirac operator.  Note that this is true even in
the vicinity of the second order phase transition point
where it has been argued that the flavor symmetry breaking
should be small due to the continuum limit being approached.
We show that at weaker couplings the flavor
splitting is drastically reduced by stout link smearing,
while this mechanism is ineffective at the stronger
couplings relevant to suspended graphene.  We also measure the average plaquette
and describe how it calls for a reinterpretation of
previous lattice Monte Carlo simulation results, due to tadpole
improvement.  After taking into account these effects, 
we conclude that previous lattice simulations
are possibly indicative of an insulating phase, although the effective
number of light flavors could be effectively less than two due to the
flavor splitting effects.  If that is true, then simulations
with truly chiral fermions (such as overlap fermions) are
needed in order to settle the question.
\end{abstract}

\maketitle

\renewcommand{\thefootnote}{\arabic{footnote}}

\section{Introduction}
Recently, a number of lattice Monte Carlo simulations of graphene
and graphene-like systems have appeared \cite{DL,DL2,Drut:2010kp,Hands}.  
Refs.~\cite{DL,DL2,Drut:2010kp} study the effective theory
of $N_f$ flavors of massless four component Dirac fermions, constrained to
2+1 dimensions, subject to an instantaneous 3+1 dimensional
Coulomb interaction \cite{eft,Son}.  In Ref.~\cite{Hands},
a 2+1 dimensional Thirring-like model is investigated.  This
is related to graphene-like systems through a large $N_f$ or strong
coupling equivalence in the dispersion relation for the auxiliary
boson versus photon.
Graphene has $N_f=2$, but
studying other $N_f$ is of interest in order to understand
the phases of such theories more generally, and because
the large $N_f$ limit is under theoretical control \cite{Son}.  Other interesting studies 
coming from the effective field theory perspective
have also recently appeared \cite{DS,Chakrabarti:2009sa}.

In this article we address the flavor symmetry breaking that
is introduced when staggered fermions are used in the lattice
formulation.  We also discuss the effect of photon tadpoles
that come from lattice field theory.  We will show that both
features play an important role in the interpretation of
lattice results.  We explore various improvements to the
lattice formulation.  One is adding a Naik term to the action,
which reduces discretization errors from $\ord{a}$ to $\ord{a^2}$,
where $a$ is the lattice spacing.  Another is tadpole improvement,
which removes ultraviolet divergent renormalizations associated with the
lattice link operators.  A final improvement that we consider
is stout link smearing, which we find restores flavor
symmetries at weak couplings but not at the strong couplings
relevant to suspended graphene.  Importantly, we find
that flavor symmetry breaking is significant in the
vicinity of the second order phase transition point
that occurs in the noncompact gauge formulation.  Thus
although it has been argued \cite{Drut:2010kp} that the
continuum limit should be approached at this point, and
hence flavor symmetry violations (which are $\ord{a}$)
should be small in this regime, we have empirical results
which contradict this expectation.  Finally, we discuss
how the flavor symmetry violations, revealed in split
eigenvalues of the Dirac operator spectrum, perhaps imply that
there are effectively less light flavors than two.  Given
the phase diagram that has been suggested by a number of
studies in the $N_f$ versus inverse coupling plane, this
would imply that the critical coupling for $N_f=2$ would
occur at a somewhat stronger coupling than is found from staggered
fermions.  Only a simulation with truly chiral lattice
fermions, such as overlap (Neuberger) fermions \cite{Neuberger:1997fp}, can conclusively answer
the question of what is the critical coupling for $N_f=2$,
since no systematic way of restoring the flavor symmetry
has been found so far for the staggered fermion formulations
at the stronger values of couplings.

The outline of this paper is as follows.  In Section \ref{conact}
we describe the action of the continuum effective theory
that is supposed to describe the low energy limit of 
suspended graphene.  We pay particular attention to
redefinitions that are involved in going to the action
in its simplest form, as these will be mirrored in redefinitions
made in the lattice formulation.  It will be shown
that in the massless limit there is only one parameter in
the theory, a coupling $g$ which is strong in the case
of suspended graphene.  We also describe the $U(4)$ flavor
symmetry of the effective theory, which is spontaneously
broken to $U(2) \times U(2)$ by the formation
of a chiral condensate, when the coupling $g$ is
sufficiently strong.  In Section \ref{slatact} we
discretize the continuum action, formulating the lattice
theory with staggered fermions.  We show the redefinitions
that isolate the one parameter of the lattice theory
(in the massless limit), $\beta=1/g^2$.  It is important
here that we make redefinitions that maintain the
unitarity of links; i.e., $U(n)= \exp[i \theta(n)]$,
where $\theta(n)$ is a real lattice field representing
the scalar potential associated with the instantaneous
Coulomb interaction.  Interestingly, this approach
demands an anisotropic lattice with lattice spacing
$a_t$ in the time direction and $a_s$ in the spatial
directions, with the anisotropy parameter $a_s/a_t$
set equal to the Fermi speed, $a_s/a_t=v_F$.  Flavor
symmetry violation of the unimproved staggered
fermion formulation is discussed in Section \ref{fsvsec}.
For 2+1-dimensional staggered fermions, $\ord{a_s,a_t}$ terms reduce the
$U(4)$ flavor symmetry to $U(1) \times U(1)$ in the massless limit.
We evaluate the spectrum
of the unimproved Dirac operator on a large number of
lattice field configurations that we have generated
by Monte Carlo techniques.  We show that at stronger
values of the coupling $g$ (equivalent to small values
of $\beta$), the flavor symmetry violation is severe.
This is revealed by the lack of four-fold spectrum
degeneracies that would be present if the $U(4)$
symmetry were respected.  We find that this is even
true near the second order phase transition point of the noncompact
gauge formulation.  

Section \ref{improve} describes
Naik fermion and tadpole improvements to the lattice formulation.
We do not find any restoration of flavor degeneracy but do
find significant reinterpretation of the bare lattice parameters
in terms of those that are tadpole improved, at strong coupling.
It will be seen that
that has important implications for the phase diagram
of the theory.  We measure the average plaquette
in dynamical simulations.  We will show that for stronger couplings the resulting tadpole improvement
of the theory has a large effect when relating the simulation lattice coupling $\beta$
to the coupling in the tadpole improved action, $\beta_{TI}$.
The result is that for the noncompact gauge action the insulator/semi-metal transition occurs
at a physical coupling that is significantly smaller than the $g^2$ of
suspended graphene.  The apparent absence of a
spectral gap in the experimental results for suspended
graphene near the Dirac K points~\cite{xp3} is in conflict
with the lattice simulations, and we will not be able
to provide an explanation for this discrepancy.

The topic of stout link smearing, which is also a type of
improvement, is discussed in Section \ref{stout}.  We find
that this is very effective at weak couplings, but that
it is not useful for restoring flavor symmetry at the
strong coupling relevant to either graphene or the
second order phase transition point that occurs in the
noncompact gauge action.  We conclude in Section \ref{conclu}
with a number of observations, summarizing our finding.

\section{Continuum action}
\label{conact}
\subsection{The effective coupling $g$}
The Euclidean spacetime action for the effective theory
is given by
\beq
S &=& \int dt d^2 x ~ \sum_{\alpha=1,2} \big( \psib_\alpha \gamma_0 D_t \psi_\alpha
+ \hbar v_F \sum_{i=1,2} \psib_\alpha \gamma_i \p_i \psi_\alpha
\ddd + m c^2 \psib_\alpha \psi_\alpha \big)
 + \frac{\e_0}{2} \int dt d^3 x ~ \sum_{i=1}^3 ~ (\p_i A_0)^2
\label{stact}
\eeq
Here $\gamma_i,$ $i=0,1,2$, are Euclidean Dirac matrices satisfying
the $SO(3)$ Euclidean rotation group
 Clifford algebra $\{ \gamma_i, \gamma_j \} = 2 \delta_{ij}$.  For instance we could choose
\beq
\gamma_i = \begin{pmatrix} 0 & i \s_i \cr
-i \s_i & 0 \end{pmatrix}, \quad i=0,1,2,
\label{npm}
\eeq
composed of Pauli matrices with $\sigma_0 \equiv \sigma_3$.
Also note that due to the nonrelativistic approximation, the covariant
derivative only involves the scalar potential $A_0$
\beq
D_t \equiv \hbar \p_t - i e A_0
\label{covar}
\eeq
  
Next we make the redefinitions
\beq
x_0 = v_F t, \quad A_0 = \frac{\hbar v_F}{e} A'_0, \quad
D_0 = \p_0 - i A'_0 
\label{cdef}
\eeq
to obtain
\beq
&& \frac{1}{\hbar} S = \int d^3 x ~ \sum_{\alpha=1,2} \big( \psib_\alpha \gamma_0 D_0 \psi_\alpha
+ \sum_{i=1,2} \psib_\alpha \gamma_i \p_i \psi_\alpha
\ddd \quad + \frac{m c^2}{\hbar v_F} \psib_\alpha \psi_\alpha \big)
 + \frac{\e_0 \hbar v_F}{2 e^2} \int d^4 x ~ \sum_{i=1}^3 ~ (\p_i A'_0)^2
\label{herf}
\eeq
Recall that the Euclidean path integral that defines
the theory has as its integrand $\exp(-S/\hbar)$.
The rescalings have isolated the sole coupling constant
in the theory,
\beq
g^2 \equiv \frac{e^2}{\hbar v_F \e_0} = (c/v_F) {4 \pi \alpha}
\label{uyuy}
\eeq
where $\alpha$ is the fine structure constant.  A final redefinition
\beq
A'_0 = g \tilde A_0, \quad \tilde D_0 = \p_0 - i g \tilde A_0
\eeq
makes it clear that $g$ is the coupling constant in the photon-electron-electron
vertex of this theory.  

Perturbation theory would be valid in the limit where
$\alpha_g \equiv \frac{g^2}{4\pi} = \alpha ~ c/v_F  \ll 1$,
which is clearly not the case for graphene, where
$c/v_F \approx 300$.  Given that the coupling is in fact strong,
it is natural to appeal to lattice Monte
Carlo methods, as has been done in the case of
the nuclear strong interaction, quantum chromodynamics (QCD).
It also becomes clear why one would like to be able
to adjust $v_F$ experimentally, since the coupling of
the theory determines the binding energy of any
possible bound states that might form from the
massless quasiparticles, analogous to hadrons in QCD.
In fact, \cite{DL,DL2,Drut:2010kp,Hands} argue that the theory is quite
similar to QCD in that when the coupling is strong enough
one creates a nonzero ``chiral'' condensate 
$\vev{\bar\psi_\alpha \psi_\beta}\not= 0$, so that the
theory is in a Mott insulator phase.  (Properly
speaking, chirality does not exist in 2+1 dimensions.  It is,
rather, a flavor symmetry that is being spontaneously broken in
the 2+1 dimensional effective theory.)

\subsection{Symmetries}
The three-dimensional $SO(3) \simeq SU(2)$ rotation group 
acting on the spinors has generators $S_{i j} = \half \sigma_{i j} \otimes \bm 1$ where
\beq
\sigma_{i j} = -(i/2) [ \gamma_i, \gamma_j] = \e_{ijk} \diag(\sigma_k, \sigma_k)
\label{s3g}
\eeq
and the $\bm 1$ factor in $\half \sigma_{i j} \otimes \bm 1$ acts on
the two dimensional flavor space.
The action \myref{npm} has a $U(4)$ flavor symmetry, with
16 generators that commute with those of the rotation group, \myref{s3g}:
\beq
&&  \bm 1 \otimes \bm 1, \quad \bm 1 \otimes \s_i, \quad
 \gamma_4 \gamma_5 \otimes \bm 1, \quad \gamma_4 \gamma_5 \otimes \s_i 
 \\
 &&\gamma_4 \otimes \bm 1, \quad \gamma_4 \otimes \s_i, \quad
 \gamma_5 \otimes \bm 1, \quad \gamma_5 \otimes \s_i 
\label{broken}
\eeq
where $\gamma_{4,5}$ are given by
\beq
\gamma_4 = \begin{pmatrix} 0 & 1 \cr 1 & 0 \end{pmatrix}, \quad
\gamma_5 = \begin{pmatrix} -1 & 0 \cr 0 & 1 \end{pmatrix}
\label{ogam}
\eeq
when we choose the Dirac matrices \myref{npm}.
A mass term $m \sum_\alpha \psib_\alpha \psi_\alpha$
reduces the symmetry to $U(2) \otimes U(2)$ since the
generators \myref{broken} are broken.  However, we still expect
a four-fold degeneracy in the spectrum of the Dirac operator
\beq
M = \gamma_0 D_0 + \sum_{i=1,2} \gamma_i \p_i + \frac{mc^2}{\hbar v_F}
\label{dirop}
\eeq
because the $\bm 4$ representation of $U(4)$ decomposes
to a $(\bm 2, \bm 2)$ representation of the subgroup $SU(2) \otimes SU(2)$.
(In spin language, this is the $(j_1,j_2)=(1/2,1/2)$ representation
of $SU(2)_1 \otimes SU(2)_2$.)  This is important in our considerations
below because the Monte Carlo simulations are done at a nonzero
mass, in order to avoid numerical difficulties (inversion of
a poorly conditioned Dirac matrix).  We will examine the spectrum
of the Dirac operator on the lattice and compare to this
four-fold degeneracy of the continuum theory with a mass term.

The formation of a ``chiral'' condensate 
$\vev{\psib_\alpha \psi_\beta} \not= 0$ in the $m \to 0$ limit
would signal a spontaneous breaking of the $U(4)$ symmetry.
In the case $\vev{\psib_\alpha \psi_\beta} \propto \delta_{\alpha \beta}$
the symmetry is reduced to $U(2) \otimes U(2)$ and in the massless
limit $m \to 0$ there will be eight massless Goldstone pseudoscalar
modes, parameterizing the coset $U(4) / U(2) \otimes U(2)$, with
a low energy dynamics described by the corresponding chiral perturbation theory.
The formation of the chiral condensate requires
a sufficiently strong value of $g$, so there is a phase boundary at
which the condensation ``turns on.'' The works \cite{DL,DL2,Drut:2010kp,Hands} have
located this phase boundary using lattice Monte Carlo methods.

\section{Discretization}
\label{slatact}
The fermionic part of the action \myref{stact} is easily discretized using the
staggered fermion formulation \cite{Kogut:1974ag}.  The gauge field part
of the action can be discretized in two ways, compact and noncompact, both
of which will be described and used here.
From this point on we work in units where $\hbar=c=1$,
and use a lattice spacing $a_t$ in the time direction
and $a_s$ in the spatial directions.  Thus we have
lattice fields at the sites $t= a_t n_0$, $x_i = a_s n_i$ ($i=1,2,3$)
where $n_0,\ldots,n_3$ are integers.  We are permitting
$a_s \not= a_t$ because the anisotropy parameter $a_s/a_t$
will provide us with the handle to remove the
Fermi velocity $v_F$ from the lattice action, so that
the only parameters that will appear are the coupling \myref{uyuy}
and the fermion mass (which must eventually be taken to zero).
This mirrors the continuum redefinition $x_0 = v_F t$ which
appears in \myref{cdef}.  The lattice action takes the form
\beq
S &=& \half \sum_{n_0 n_1 n_2} a_t a_s^2 \bigg\{ \frac{1}{a_t}
( \bar \chi(n) U(n) \chi(n+ \hat 0) - \bar \chi(n) U(n-\hat 0) \chi(n-\hat 0) )
\ddd + v_F \frac{1}{a_s} \sum_{i=1}^2 \eta_i(n) ( \bar\chi(n) \chi(n+\hat \imath)
- \bar\chi(n) \chi(n-\hat \imath) ) + m \bar \chi(n) \chi(n) \bigg\}
\ddd + \sum_{n_0 \ldots n_3} a_t a_s^3 \frac{\e_0}{2} 
\sum_{i=1}^3 \( \frac{\theta(n) - \theta(n-\hat \imath)}{a_s} \)^2
\eeq
The notation employs four-vectors $n=(n_0,n_1,n_2,n_3)$ and
unit vectors $\hat 0 = (1,0,0,0)$ etc.
Here $\chi,\chib$ are 1-component fermions and as site dependent coefficients one has
the staggered phase factors $\eta_1(n) = (-1)^{n_0}$ and $\eta_2(n) = (-1)^{n_0 + n_1}$.
The reason that one-component fermions can be used is because 
staggered fermions ``suffer'' from doubling, so that in three dimensions there are eight continuum
modes, which organize themselves into two four-component fermions under a change
of basis.\footnote{This is analogous to the four
flavors that appear in the 3+1 dimensional staggered formulation of lattice quantum
chromodynamics.}  The link fields are defined as $U(n) = \exp(i e a_t \theta(n))$,
where $\theta(n)$ is
the lattice version of the scalar potential $A_0(x)$.  Here we have used
the noncompact form of the gauge action in the last term.  The compact form
will be discussed at a later point below.

We next rescale to dimensionless fields, $\chi \to \chi/a_s$ and
$\theta \to \theta / a_t e$ to obtain:
\beq
S &=& \half \sum_{n_0 n_1 n_2} \bigg\{ 
\bar \chi(n) U(n) \chi(n+ \hat 0) - \bar \chi(n) U(n-\hat 0) \chi(n-\hat 0) 
\ddd + v_F \frac{a_t}{a_s} \sum_{i=1}^2 \eta_i(n) ( \bar\chi(n) \chi(n+\hat \imath)
- \bar\chi(n) \chi(n-\hat \imath) ) + m a_t \bar \chi(n) \chi(n) \bigg\}
\ddd 
+ \sum_{n_0 \ldots n_3} \frac{a_s}{a_t} \frac{\e_0}{2} 
\sum_{i=1}^3 ( \theta(n) - \theta(n-\hat \imath) )^2
\eeq
Finally, we can absorb the Fermi speed $v_F$ into the
anisotropy parameter, choosing $a_s/a_t = v_F$, to obtain the
lattice action in its most convenient form,
\beq
S &=& \half \sum_{n_0 n_1 n_2} \bigg\{ 
 \bar \chi(n) U(n) \chi(n+ \hat 0) - \bar \chi(n) U(n-\hat 0) \chi(n-\hat 0) 
\ddd + \sum_{i=1}^2 \eta_i(n) ( \bar\chi(n) \chi(n+\hat \imath)
- \bar\chi(n) \chi(n-\hat \imath) ) + \hat m  \bar \chi(n) \chi(n) \bigg\}
\ddd
+ \sum_{n_0 \ldots n_3} \frac{\beta}{2} 
\sum_{i=1}^3 ( \theta(n) - \theta(n-\hat \imath) )^2
\label{latar}
\eeq
where
\beq
\beta = \frac{1}{g^2} = \frac{v_F \e_0}{e^2}, \quad \hat m = m a_t.
\eeq
A slightly different choice for the anisotropy parameter
$a_s/a_t$ will be made below when we come to tadpole improvement.

We also consider the case of a compact gauge action,
where the last term in \myref{latar} is replaced
by
\beq
- \beta \sum_{n_0 \ldots n_3} \sum_{i=1}^3 \real U(n) U^*(n-\hat\imath)
\label{coga}
\eeq
In the weak field limit (small $\theta(n)$), which corresponds
to large $\beta$, the two formulations are equivalent.  However,
at small $\beta$ it is expected that there will be qualitative
differences.

\section{Flavor symmetry violation}
\label{fsvsec}
As stated above, a single staggered fermion
automatically yields two flavors, since the 
staggered formulation does not fully solve the doubling problem.
In the continuum, the massless theory with two flavors
has a $U(4)$ flavor symmetry, which is reflected in
a degeneracy of the spectrum of the Dirac operator.
On the other hand, it
is known that the leading order spectral degeneracies
of the lattice Dirac operator are broken by flavor 
violating higher order terms (in the lattice spacings $a_t,a_s$).   
In the massless limit but at nonzero lattice spacing only
a $U(1) \otimes U(1)$ flavor symmetry remains
(in addition to some discrete symmetries) \cite{Golterman:1984cy}.
Long ago it was shown in the 3+1 dimensional case that
the flavor symmetry breaking can be seen by going to the ``flavor basis'' \cite{KlubergStern:1983dg}.
For 2+1 dimensions, see for 
example \cite{DelDebbio:1997dv} where staggered fermion
flavor breaking terms were previously considered in the context of
the Thirring model.  Thus in the present article
we are reiterating concerns that were raised already in \cite{DelDebbio:1997dv},
though here our principal concern is the effect in the context of graphene effective 
lattice field theory.  Although the flavor symmetry breaking terms
are irrelevant operators (i.e., they are suppressed by $a_t,a_s$),
at one loop and at finite lattice
spacing they have important effects on the self 
energy of the fermions \cite{Golterman:1984cy}.  
The effect of this
flavor symmetry violation on the order parameter $\vev{\psib \psi}$
that is used to distinguish the semi-metal versus insulator
phases is not known, though in our Conclusions
we will make a conjecture for what might occur.  
The flavor changing interactions are a lattice artifact
that is known to disappear in the continuum limit.
Hence, if one could send the lattice spacings $a_t,a_s$ of the
discretized effective theory (not to be confused with the lattice
constant of the graphene system itself) to zero, one
would recover the full $U(4)$ symmetry \cite{Golterman:1984cy}.
However, the Monte Carlo simulations are performed at
finite $a_t,a_s$, and so this lattice artifact must be taken
into account.  Thus it is not quite accurate
to say that one is simulating the effective theory
with two (1+3)-dimensional Dirac fermions constrained
to a plane, equivalent to four massless (1+2)-dimensional
Dirac fermions.  An extrapolation in the lattice spacing
or suppression of the lattice artifacts is needed.
One would like a systematic way to remove these lattice artifacts.  
This motivates the present study.

We determine the size of the flavor-splitting
by studying the eigenvalues of the lattice Dirac operator,
which is the discretization of \myref{dirop} corresponding
to the lattice action \myref{latar}.
In Fig.~\ref{compare_small} the ``unimproved'' data
shows the average spectrum of the staggered Dirac operator,
for the lowest lying modes.  Here a Monte Carlo
simulation was performed with $\beta=0.11$, and
eigenvalues were obtained for each configuration
of the gauge field.  The error bars in the figure indicate the standard deviation in the
eigenvalues.  It can be seen that there
is a linear rise in eigenvalues, with no degeneracies
whatsoever.  Thus at strong coupling the flavor
symmetry of the continuum is badly broken.

Next we consider the case of weak coupling, $\beta=4.0$.
In Fig.~\ref{compare_big} the unimproved data
does show evidence of approximate degeneracies.  The
weaker coupling leads to smoother configurations of the gauge field.
Rough gauge fields are farther away from the continuum
limit, so that the $\ord{a_t,a_s}$ flavor
symmetry violations is more pronounced.

We have examined the spectrum for other values of $\beta$.
The general pattern is that for strong coupling the flavor
symmetry is badly broken.  Our next task is to
attempt to restore it, since the $\beta$ corresponding
to graphene and the phase transition of the effective
theory is at a strong coupling value.

\section{Improvement}
\label{improve}
In fact, some time ago the lattice QCD community set aside
unimproved staggered fermions due to unwanted lattice
artifacts.  Modern staggered fermions are improved in
various ways in order to suppress these effects \cite{Davies:2003ik,ptart}.  So-called
AsqTad staggered fermions were popular for several
years for the study of K and B physics (e.g.~\cite{Bernard:1997mz}).  Further improvements
have been introduced to produce HISQ staggered fermions \cite{Follana:2006rc}.
Detailed studies of the low lying eigenvalue spectrum of
various staggered Dirac operators have for instance been
conducted in \cite{Durr:2004as}.
In each case, an important effect is to restore the flavor
degeneracy by suppressing flavor changing interactions.
The present work represents a first attempt in that direction; however,
we will find that improvement of staggered fermions
in the present context is more difficult.  The reason
is that for the study of graphene and the
phase transition of the effective theory
the coupling is strong, where the flavor symmetry
is badly broken.

In lattice QCD it is known that flavor symmetry breaking can be ameliorated by making
improvements to the lattice action that reduce lattice artifacts.  An expansion in the lattice spacing
$a$ (or $a_t,a_s$ in our case)
and gauge coupling $g$ allows for coefficients of various improvement terms
to be determined in perturbation theory.  However, asymptotic freedom
should be important, since in that case it is clear how one makes
these coefficients small in matching onto the desired continuum theory.
It is then an important question
whether for the strongly coupled theory of graphene,
where there is no asymptotic freedom,
the lattice action can be improved so as to
reduce the flavor symmetry breaking effects.
Certainly perturbative improvement is out of the question.

\subsection{Tadpole improvement}
Tadpoles arise from $\vev{A_0^2(x)} 
\sim \vev{\theta^2(n)} \sim 1/a_t^2$, where the
estimate is made on dimensional grounds.
As mentioned above, 
we study both the compact and noncompact gauge actions.
In the noncompact case, gauge field tadpoles
only enter the perturbation series through
the gauge links $U(n)=\exp(i a_t e \theta(n))$
that are contained within the fermion action.  In
the compact case there are additional multiphoton
vertices coming from expansion of the gauge
action \myref{coga}.  Consider the following example in the
fermion time-like hopping terms.  In this, we reintroduce
dimensions and canonical kinetic term for $\theta(n)$ through $\theta(n) \to a_t g \theta(n)$.
Then expanding the link $U(n)=\exp[i a_t g \theta(n)]$ and focusing
on the contribution to the fermion self-energy,
we obtain a term $a_t g^2 \vev{\theta^2(n)} \chib(n) \chi(n + \hat 0)
\sim (g^2/a_t) \chib(n) \chi(n + \hat 0)$.  I.e., there is a
large correction to the hopping term, even though the $\theta^2 \chib \chi$
vertex is irrelevant by power counting.
There is also a large effect on the marginal $\theta \chib \chi$ vertex:
\beq
&& i g \theta(n) \( 1 - \half a_t^2 g^2 \vev{\theta^2(n)} + \cdots \) 
\chib(n) \chi(n + \hat 0) \nnn
&& = i g \theta(n) \( 1 + \ord{g^2} \) \chib(n) \chi(n + \hat 0)
\eeq
Here again, the correction is $\ord{g^2}$ rather than
$\ord{g^2 a_t^2}$, due to the tadpole $\vev{\theta^2(n)} \sim 1/a_t^2$.
The tadpoles associated with the irrelevant fermion vertices
thus give significant contributions to the renormalization
of $g$, causing the matching onto continuum perturbation
theory to be problematic.  This can be circumvented
through a change in renormalization scheme, known as 
tadpole improvement \cite{Lepage:1992xa}.
In fact, since for graphene
the value of $g$ is large, the tadpoles corrections are out
of perturbative control and must be evaluated
nonperturbatively.  

We will now show that the translation between the bare lattice
$\beta=1/g^2$ (i.e., the parameter that appears in the action
that is simulated) and its tadpole improved value $\beta_{\text{TI}}$ is
somewhat different depending on whether the compact or noncompact
form of the gauge action is used, more so at stronger values of the coupling.
From this, the physical coupling---as estimated by the
tadpole improved value $\beta_{\text{TI}}$---is different from the bare coupling,
due to radiative effects.  In fact, we will reproduce the
results of \cite{Drut:2010kp} regarding the relationships
between $\beta_{\text{TI}}$ and $\beta$.

We begin with the expectation value $\vev{P}$
of the plaquette operator $P=U(n) U^*(n+\hat \imath), \;
i=1$ or $2$, which is related to $\vev{A_0^2(x)}$.
The average link $u_0$ is defined through this quantity:
\beq
u_0 = \vev{P}^{1/2}
\label{u0d}
\eeq
Note that the square
root is used here, in contrast to the fourth root that appears in
QCD applications, since the plaquette operator is quadratic in the links
that are allowed to fluctuate in the present, nonrelativistic 
formulation.

Tadpole improvement \cite{Lepage:1992xa}
can be understood as integrating out ultraviolet
modes of the link operator $U(x)$, to obtain an
effective infrared link operator.
The quantity $u_0$ represents the ultraviolet divergent
effects of tadpoles $\vev{A_0^2}$.
Thus, the link is related to an infrared (IR) field
$V(n)$ or $\theta^{IR}(n)$:
\beq
U(n) \equiv u_0 V(n) \approx u_0 ( 1 + i a_t e \theta^{IR}(n) )
\eeq

When the lattice is formulated using instead the $V(n)=U(n)/u_0$ links,
one has
\beq
S &=& \sum_{n_0 n_1 n_2} \bigg\{ 
\frac{a_s^2}{u_0} ( \bar \chi(n) U(n) \chi(n+ \hat 0) - \bar \chi(n) U(n-\hat 0) \chi(n-\hat 0) )
\ddd + v_F a_t a_s \sum_{i=1}^2 \eta_i(n) ( \bar\chi(n) \chi(n+\hat \imath)
- \bar\chi(n) \chi(n-\hat \imath) ) + m a_t a_s^2 \bar \chi(n) \chi(n) \bigg\}
\ddd + \sum_{n_0 \ldots n_3} a_t a_s \frac{\e_0}{2} 
\sum_{i=1}^3 ( \theta(n) - \theta(n-\hat \imath) )^2
\eeq
The redefinition of variables is now
\beq
\chi = \frac{\sqrt{u_0}}{a_s} \chi', \quad \theta = \frac{1}{a_t e} \theta'
\label{rewy}
\eeq
One finds that $a_s/a_t = v_F u_0$ simplifies the spatial
derivative term and that the result is equation \myref{latar} except that
$\beta$ and $\hat m$ are replaced by
\beq
\beta = u_0 \frac{v_F \e_0}{e^2} = u_0 \beta_{\text{TI}}^{\text{nc}}, 
\quad \hat m = u_0 \hat m_{\text{TI}}
\label{tinc}
\eeq
Note that $\beta_{\text{TI}}^{\text{nc}}$ 
and $\hat m_{\text{TI}}$ are what would have appeared in
the lattice action had we not included $u_0$ in the redefinition
\myref{rewy}.  Hence these are the inverse coupling and dimensionless mass of the
tadpole improved action.  By contrast, $\beta$ and $\hat m$ are the
inverse coupling and mass that are used in the simulation after
going to the redefined variables where the action
takes its simplest form (i.e., $u_0$ does not appear
explicitly).
Thus in the massless limit, for the noncompact gauge action,
the entire effect of the tadpole improvement is to rescale
the inverse coupling according to this equation.  Something
similar occurs in the compact gauge
action case.  There we have in addition a factor $1/u_0^2$
in front of the gauge term,
\beq
\sum_{n_0 \ldots n_3} \frac{1}{u_0^2} \frac{a_s}{a_t} \frac{\e_0}{2 e^2}
\sum_{i=1}^3 U(n) U^*(n+\hat i)
\eeq
Here then the result is
\beq
\beta = \frac{1}{u_0} \beta_{TI}^c
\label{ticom}
\eeq
These rescalings of $\beta$ agree with those found recently
in \cite{Drut:2010kp}.

\subsection{Naik improvement}
The Naik \cite{Naik:1986bn} fermion action improvement reduces discretization
errors and when the tadpole improvement is also performed it is given by:
\beq
S_N &=& a_s^2 \sum_{n_0 n_1 n_2} \chib(n) \frac{1}{2} \bigg\{
\frac{c_1}{u_0} [ U(n) \chi(n+\hat 0) - U^*(n-\hat 0) \chi(n-\hat 0)] 
\ddd + \frac{c_2}{u_0^3} [U(x) U(n+\hat 0) U(n+2\hat 0) \chi(n+3 \hat 0)
\ddd - U^*(n-\hat 0) U^*(n-2 \hat 0) U^* (n-3 \hat 0) \chi(n-3 \hat 0)] \bigg\}
\ddd + v_F a_s a_t \sum_{i,n_0 n_1 n_2} \eta_i(n) \chib(n) \frac{1}{2} \bigg\{
c_1 [ \chi(n+\hat \imath) - \chi(n-\hat \imath)] 
+ c_2 [ \chi(n+3 \hat \imath)
- \chi(n-3\hat \imath)] \bigg\}
\ddd + a_s^2 a_t m \sum_{n_0 n_1 n_2} \chib(n) \chi(n)
\label{naika}
\eeq
Tree level improvement makes the action $\ord{a^2}$ accurate
by setting $c_1=9/8, \; c_2 = -1/24$.

Next we make the redefinitions \myref{rewy}, together with setting
$a_s/a_t = v_F u_0$ as before, to obtain:
\beq
S_N &=& \sum_{n_0 n_1 n_2} \chib'(n) \frac{1}{2} \bigg\{
c_1 [ U(n) \chi'(n+\hat 0) - U^*(n-\hat 0) \chi'(n-\hat 0)] 
\ddd + \frac{c_2}{u_0^2} [U(x) U(n+\hat 0) U(n+2~\hat 0) \chi'(n+3~ \hat 0)
\ddd - U^*(n-\hat 0) U^*(n-2 ~\hat 0) U^* (n-3~ \hat 0) \chi'(n-3 ~\hat 0)] \bigg\}
\ddd +  \sum_{i,n_0 n_1 n_2} \eta_i(n) \chib'(n) \frac{1}{2} \bigg\{
c_1 [ \chi'(n+\hat \imath) - \chi(n-\hat \imath)] 
+ c_2 [ \chi'(n+3 \hat \imath)
- \chi'(n-3\hat \imath)] \bigg\}
\ddd + \hat m \sum_{n_0 n_1 n_2} \chib'(n) \chi'(n)
\label{naikb}
\eeq
where again, $\hat m = m a_t u_0$.
This is the ``Naik-tadpole improvement;'' note that $u_0$ appears
explicitly in this action.
To obtain just the Naik improvement, one can set $u_0=1$ in the
previous expressions.

\subsection{Spectrum results}
\begin{figure}
\begin{center}
\includegraphics[width=2.2in,height=2.5in,angle=90]{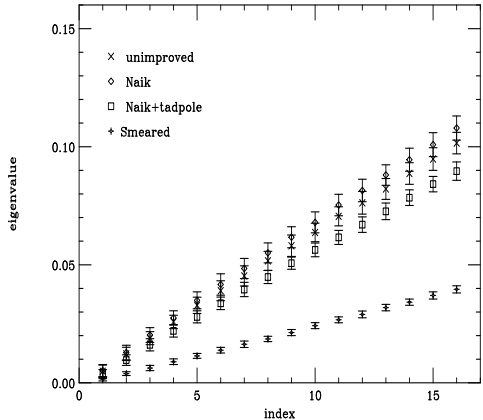}
\caption{Spectrum of lowest lying modes of the three massless Dirac operators
we consider, for compact gauge action. The configurations of gauge fields were dynamically
generated at $\beta=0.11$ and $m=0.01$ on a $12^3 \times 8$ lattice
with the unimproved staggered fermion action and plaquette gauge term.
The tadpole improvement of the Naik Dirac operators 
used $u_0=0.256$.  Average eigenvalues are shown, and the
error bars represent standard deviations.  \label{compare_small}}
\end{center}
\end{figure}

\begin{figure}
\begin{center}
\includegraphics[width=2.2in,height=2.5in,angle=90]{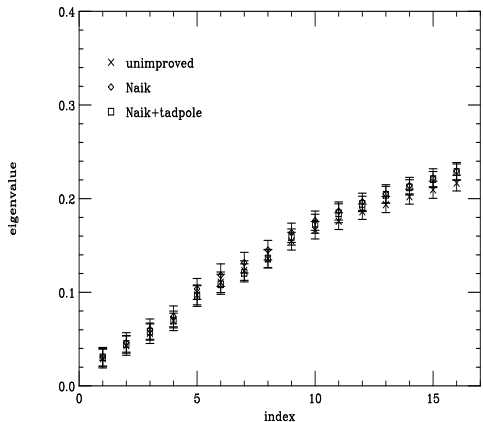}
\caption{Same as Fig.~\ref{compare_small} except that here we use
the noncompact gauge action.  \label{compare_small_nc}}
\end{center}
\end{figure}

We have computed the low lying eigenvalues of the spectrum of the Dirac operator 
on dynamical configurations at various values of $\beta$,
in order to see the size of the flavor symmetry violating effect.
Fig.~\ref{compare_small} shows the spectrum of average eigenvalues
for $\beta=0.11$, $\hat m=0.01$ on $12^3 \times 8$ lattices, 
with compact gauge action, as well as the standard deviation (by error bars).
Fig.~\ref{compare_small_nc} shows the same thing except that the
noncompact gauge action was used.  In either case,
one can see that there is no hint of the four-fold degeneracy of the continuum
theory and that the splitting is of the order $0.02$.  
By comparison, the explicit mass 
in the simulations of \cite{DL} ranged from 0.0025 to 0.02.
Thus the flavor changing interactions split the spectrum at the order of
the mass or greater, and one is far from the
desired theory.  Since according the to Banks-Casher 
relation \cite{Banks:1979yr}
the condensate on the lattice is determined by
the density of near-zero modes, a significant systematic
error will be introduced by the flavor splitting that we observe.
We note that for the ``improved'' Dirac operators the splitting
is not at all improved.  This would seem to indicate
that the lattice is actually quite coarse, so that
suppressing lattice artifacts cannot be achieved
by simple power-counting in the lattice spacings $a_t,a_s$,
such as is done in the Naik improvement.  It is also
worth mentioning that large scaling violations were
seen in \cite{DL} for strong coupling (very small values
of $\beta$) which would be a further indication that
lattice artifacts are playing a dominant role.
However, the fact that \cite{DL} observe scaling in
a regime where we see large flavor violations is
interesting, as it suggests that there is a universal
description but that it is one with less flavor
symmetry than the $U(4)$ of the target graphene effective
theory.

\begin{figure}
\begin{center}
\includegraphics[width=2.2in,height=2.5in,angle=90]{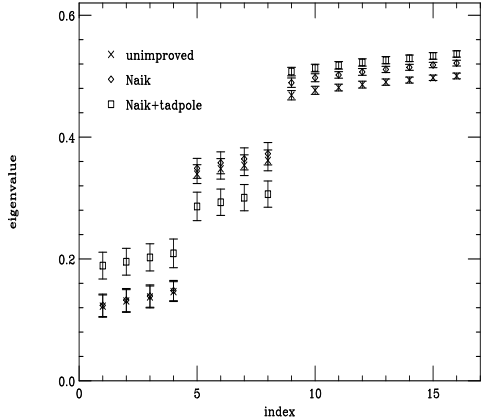}
\caption{Similar to Fig.~\ref{compare_small} (compact gauge action) except that 
$\beta=4.0$ and $u_0=0.974$. \label{compare_big}}
\end{center}
\end{figure}

\begin{figure}
\begin{center}
\includegraphics[width=2.2in,height=2.5in,angle=90]{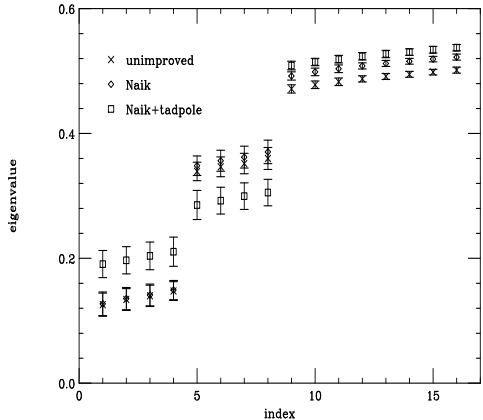}
\caption{Similar to Fig.~\ref{compare_small_nc} (noncompact gauge action) except that 
$\beta=4.0$ and $u_0=0.974$. \label{compare_big_nc}}
\end{center}
\end{figure}

As a further check, we have also computed the spectrum
from a simulation at the weak coupling $\beta=4$
where the flavor violation is expected to be small due to weak interactions.
We also note that at this weak value of the coupling the compact
and noncompact formulations of the gauge action are completely
equivalent.  Thus the flavor symmetry breaking that we next
describe is universal.
At large $\beta$ the fluctuations in the gauge field strength
are suppressed and a perturbative expansion of the link
operators $U_0(x) \approx 1 + i a g A_\mu(x)$ should be valid.
Results for the low lying eigenvalues of the three
types of Dirac operators are shown in Figs.~\ref{compare_big} and \ref{compare_big_nc},
and these certainly show a closer approximation to the four-fold degeneracy.
However, the improved Dirac operators do not show any
superiority to the unimproved one.  This somewhat surprising
result suggests that a further improvement may be needed,
such as smearing, something which we explore in
the next section.

It is also interesting to have a statistical measure for what happens to
flavor symmetry over an ensemble.  For this purpose we have computed
\beq
R_{\text{FSV}} = \frac{\bar \lambda_4 - \bar \lambda_1}
{\frac{1}{4} ( \bar \lambda_1 + \bar \lambda_2 + \bar \lambda_3 + \bar \lambda_4 ) }
\eeq
where $\bar \lambda_i$ is the average value of the $i$th eigenvalue.
This measures the relative flavor symmetry breaking in the first
four eigenvalues.  On the $\beta=0.11$ (compact) lattice we obtain $R_{\text{FSV}} \approx 1.4(1)$.
On the $\beta=4.0$ (compact) lattice we obtain $R_{\text{FSV}} \approx 0.18(2)$.
These results are independent of the improvement, which is curious
at the larger $\beta$.

\subsection{Relation between $\beta$'s}
Above, we found that the value of $\beta_{TI}$ in the tadpole
improved action can be related to another value $\beta$
obtained after redefinitions, given by Eqs.~\myref{tinc} 
and \myref{ticom}.  The latter should be used
in the simulation with an action that is equivalent to
one without tadpole improvement (or only a factor $1/u_0^2$
on the temporal Naik term).  We are therefore interested in the
effective value $\beta_{TI}$ as a function of $\beta$
so that we know how to interpret simulations done at $\beta$
in terms of the underlying $\beta_{TI}$.  For instance,
Drut and L\"ahde find a critical value of the coupling
for which a condensate forms, and this should be interpreted
as a value of $\beta$ at which the simulation is
done (i.e., in an action without $u_0$ appearing explicitly).
To see what this physically corresponds to, one must
translate back to $\beta_{TI}^{nc}$
in order to find the value of the coupling in the
tadpole improved action, where ultraviolet artifacts
are minimized.

\begin{table}
\begin{center}
\begin{tabular}{|c|c|c|c|}
\hline $\beta$ & $\vev{P}$ & $u_0$ & $\beta_{\text{TI}}^{\text{c}}$ \\ \hline
0.037 & 0.0306(40) & 0.175(11) & 0.00647(42) \\
0.058 & 0.034(3) & 0.183(7) & 0.01061(41) \\
0.11 & 0.066(4) & 0.256(8) & 0.02816(88) \\
0.15 & 0.0901(44) & 0.3002(73) & 0.0450(11) \\
0.25 & 0.1492(43) & 0.3863(56) & 0.0965(15) \\
0.5 & 0.504(6) & 0.710(4) & 0.355(2) \\
1.0 & 0.814(3) & 0.9023(14) & 0.9023(14) \\
2.0 & 0.9120(13) & 0.9550(7) & 1.91(14) \\
4.0 & 0.949(2) & 0.974(1) & 3.896(4) \\
\hline
\end{tabular}
\caption{The average plaquette $\vev{P}$ and
the tadpole correction factor $u_0$ that is derived from
it, as a function of $\beta$, for the compact gauge action.
This then gives a value for tadpole improved inverse coupling $\beta=\beta_{\text{TI}}^{\text{c}}$.
For instance, for graphene we want $\beta_{\text{TI}}^{\text{c}}=0.037$ and the
inverse coupling that should be used in the simulation is $\beta \approx 0.12$.
\label{t_trend}}
\end{center}
\end{table}

\begin{table}
\begin{center}
\begin{tabular}{|c|c|c|c|}
\hline $\beta$ & $\vev{P}$ & $u_0$ & $\beta_{\text{TI}}^{\text{nc}}$ \\ \hline
0.002 & 0.0131(37) & 0.114(16)  &  0.0175(25) \\
0.004 & 0.0108(54) & 0.104(26)  &  0.0385(96) \\
0.005 & 0.0121(41) & 0.110(19)  &  0.0455(78) \\
0.01  & 0.0121(27) & 0.110(12)  &  0.091(10)  \\
0.02  & 0.0118(42) & 0.108(19)  &  0.184(33)  \\
0.037 & 0.0272(46) & 0.165(14)  &  0.224(19) \\
0.058 & 0.0757(40) & 0.2751(73) &  0.2108(56) \\
0.11 & 0.2392(45) & 0.4891(46)  &  0.2249(21) \\
0.25 & 0.5228(40) & 0.7230(27)  &  0.3458(13) \\
0.5 & 0.7192(29) & 0.8481(17)   &  0.5896(12) \\
1.0 & 0.8466(20) & 0.9201(11)   &  1.0868(13) \\
2.0 & 0.9195(11) & 0.9589(6)    &  2.0857(13) \\
4.0 & 0.94(2)    & 0.97(1)      &  4.124(43)    \\
\hline
\end{tabular}
\caption{The average plaquette $\vev{P}$ and
the tadpole correction factor $u_0$ that is derived from
it, as a function of $\beta$, for the noncompact gauge action.
This then gives
a value for tadpole improved inverse coupling $\beta=\beta_{\text{TI}}^{\text{nc}}$.
For instance, for graphene $\beta_{\text{TI}}^{\text{nc}}=0.037$ and the
inverse coupling that should be used in the simulation is $\beta \approx 0.004$.
\label{t_trend_noncompact}}
\end{center}
\end{table}

Results for the compact action are summarized in Table \ref{t_trend}
and for the noncompact action in Table \ref{t_trend_noncompact}. 
Thus to simulate graphene, which has $\beta_{TI} \approx 0.037$,
we should choose the modified values $\beta$ given
in the first rows of Tables \ref{t_trend} or \ref{t_trend_noncompact},
depending on the form of the gauge action.  This gives $\beta \approx 0.12$
for compact and $\beta \approx 0.004$ for noncompact.
The simulation coupling where Drut and L\"ahde have found a phase
transition is $\beta_c \approx 0.074$.
The physical value of the inverse coupling is then approximately
$\beta_{TI}^{nc} \approx 0.21$, which is at a coupling significantly weaker
than graphene, $\beta_{TI} \approx 0.037$.
Thus the appearance of the condensate $\vev{\bar\psi \psi}$
occurs for a weaker value of the coupling, and will
persist at the stronger value of graphene.  One concludes
that the lattice simulation is indicative of an insulator
phase.  This is in agreement with the findings of \cite{Drut:2010kp}.

We also mention in passing
that the value of $\vev{P}$ and hence $u_0$ turned out to be essentially
independent of which fermion action (unimproved,
Naik improved or Naik-tadpole improved) we used in the
simulation.  We also changed the mass to 0.02 and find
the same value of $u_0$.


\section{Stout link smearing}
\label{stout}
We have seen that at weak coupling (large $\beta$), the
spectrum degeneracies start to appear.  This is the result
of the fact that in this regime the gauge fields are
smooth, whereas at strong coupling the gauge fields are rough.
Clearly what is needed at strong coupling is 
a way to smooth out the short distance (unphysical) roughness without
destroying the long distance (physical) fluctuations of the gauge
field.  The way that this
can be done is to use smeared links in the fermion action.
Here we will study stout link smearing \cite{Morningstar:2003gk} and will find that
it successfully restores the level degeneracies for
moderate to weak coupling, but that it fails at couplings
as strong as graphene, $\beta_{TI} = 0.037$.

Stout link smearing in the present context introduces the
definitions
\beq
C(n) = \rho \sum_{i=1}^3 [ U(n+\hat\imath) + U(n-\hat\imath) ], \quad
\Omega(n) = C(n) U^*(n), \quad Q(x) = \frac{i}{2} [ \Omega^*(n) - 
\Omega(n) ]
\eeq
and $U^{(k)}(n)$ at smearing step $k$ are mapped into
$U^{(k+1)}(n)$ according to
\beq
U^{(k+1)}(n) = \exp [ i Q^{(k)}(n) ] U^{(k)}(n)
\eeq

\begin{figure}
\begin{center}
\includegraphics[width=2.2in,height=2.5in,angle=90]{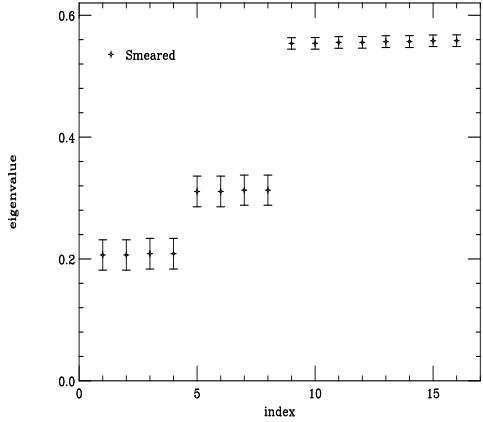}
\caption{$\beta=4.0$, compact gauge action, 10 stout link smearings
with $\rho=1/6$. \label{smear_big}}
\end{center}
\end{figure}

It can be seen in Fig.~\ref{smear_big} that smearing works
very well at weak coupling.  The smeared eigenvalue data
has 10 smearing iterations with smearing parameter $\rho=1/6$,
where the latter was found to be optimal based on trial
and error.  Less smearing iterations obviously
results in less degeneracy.  Unfortunately, as the coupling is made stronger,
the smearing becomes progressively less effective, as
can be seen in Fig.~\ref{compare_small}.

\begin{figure}
\begin{center}
\includegraphics[width=2.2in,height=2.5in,angle=90]{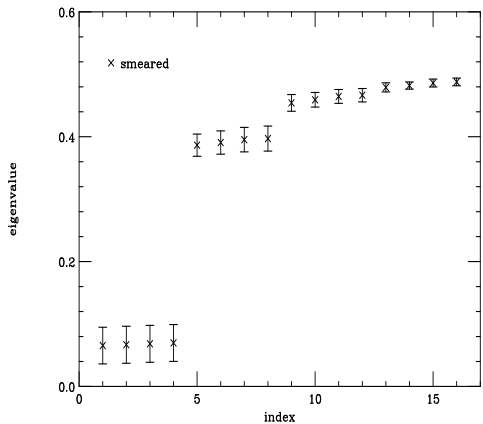}
\caption{$\beta=0.11$, noncompact gauge action, 10 stout link smearings
with $\rho=1/6$. \label{smear_small_nc}}
\end{center}
\end{figure}

By contrast for the noncompact gauge action, even at
the relatively small value of $\beta=0.11$, one finds
a significant improvement from smearing; see Fig.~\ref{smear_small_nc}.
Since the phase transition occurs at $\beta \approx 0.07$ we
expect smearing to be quite useful for reducing flavor
symmetry breaking in the vicinity of this point.  On the
other hand from Table \ref{t_trend_noncompact} we found that graphene
with $\beta_{TI}^{nc}=0.037$ corresponds to $\beta \approx 0.004$
which is far too strong for smearing to help.
Indeed we have found that there is no restoration of degeneracy
in this case.

\section{Conclusions}
\label{conclu}
We have found that at $\beta \lappeq 1$ both the unimproved action, and the
tadpole-improved action with a Naik term show significant
flavor symmetry breaking.  We have also measured the average plaquette
term used for tadpole improvement and have
described how it calls for a reinterpretation of
previous lattice simulation results.  Importantly,
it indicates that the insulator/semi-metal phase
transition observed on the lattice occurs
at a physical coupling that is significantly weaker
to the one that appears in suspended graphene.
It follows that the lattice simulations predict
that the chiral symmetry is spontaneously broken
and that suspended graphene would be in the insulating phase.

On the other hand, conjectured phase diagrams
in the $g$ versus $N_f$ plane would indicate that
the critical $g$ decreases as $N_f$ is decreased.
So, if the staggered formulation really simulates
effectively less that $N_f=2$ due to the flavor
symmetry breaking, the lattice simulations would
predict a critical $g$ that is weaker than that
of graphene.  Restoration of the $U(2N_f)$ flavor
symmetry would tend to increase the value of the
critical $g$.  Thus it is still possible, though unlikely, that lattice
simulations would predict that suspended graphene
is in the semi-metal phase, provided the full
flavor symmetry is intact.  We think that it is
unlikely since the critical $\beta_c$ would have
to shift all the way from $\beta=0.07$ to $\beta=0.004$
as a result of restoring the flavor symmetries.
Still, a study with overlap fermions is of interest
to settle the question.

We have conducted studies with both the compact
and noncompact formulation in their gauge action.
In 1+3 dimensional quantum electrodynamics,
the compact formulation has difficulties with
a bulk phase transition in the strong coupling regime,
separating it from the continuum theory (see for example \cite{Kogut:1987kg} and recent
work in \cite{Kogut:2002vu}).  On the other
hand, with the nonrelativistic constraint $U_i(x) \equiv 1, \;
i=1,2,3$ that we impose, the phase structure of the compact
theory will be quite different since, for instance, magnetic monopoles
will not exist.  However, the presence of vortices requires further
investigations of the compact theory, which we will leave
to future work.  At present what is known from \cite{Drut:2010kp}  
is that the compact theory has a first order phase
transition in contrast to the second order transition
of the noncompact case.  This very different
phase structure indicates that nonperturbative
features, such as vortices, are having a significant
effect in distinguishing the two theories at
strong coupling.  In the present article we show results
for both compact and noncompact gauge action.  We find that the qualitative
features do not change: the large flavor violations are present in
either formulation at strong coupling.

\section*{Acknowledgements}
We are grateful to Joaquin Drut for numerous insightful questions and
useful comments.  Helpful communications were also received from Simon Hands
and Timo L\"ahde.  JG was supported by Rensselaer faculty development funds
and by the U.S.~Department of Energy, Office of Science,
Office of High Energy Physics, Outstanding Junior Investigator
program, contract DE-FG02-08ER41575.
AS and SN received support for this project from the New York State 
Interconnect Focus Center at Rensselaer.

\appendix

\section{Simulation details}
All of our results were obtained using hybrid
Monte Carlo simulations with dynamical staggered fermions.
This simulation method has been reviewed in
the present context in \cite{DL2}.
The mass in our simulations
was $ma=0.01$, where $a$ is the lattice spacing.
We have simulated on various sizes of lattices ($6^3 \times 8$, $8^3 \times 8$, 
$12^3 \times 8$, $16^3 \times 8$ and $24^3 \times 8$).
We checked that the configurations
were fully thermalized by comparing ordered and disordered
starts.

\end{document}